\documentclass[12pt,a4paper]{article}
\usepackage{amsmath,amssymb}
\usepackage{graphicx,color}
\numberwithin{equation}{section} \oddsidemargin 0 mm \evensidemargin
0 mm \topmargin -10 mm \textheight 215 mm \textwidth 163 mm

\renewcommand{\thefootnote}{\fnsymbol{footnote}}
\newcommand{\nn}{\nonumber}
\begin{document}
\vspace{12mm}

\begin{center}
{{{\Large {\bf Stability of the massive graviton \\
around a BTZ black hole in three dimensions}}}}\\[10mm]

{Taeyoon Moon\footnote{e-mail address: tymoon@sogang.ac.kr} and  Yun
Soo Myung\footnote{e-mail address: ysmyung@inje.ac.kr},
}\\[8mm]

{Institute of Basic Sciences and Department of Computer Simulation, Inje University, Gimhae 621-749, Korea}\\[0pt]

\end{center}
\vspace{2mm}

\begin{abstract}
We investigate the massive graviton stability of the BTZ black hole
obtained from three dimensional massive gravities which  are
classified into the parity-even and parity-odd gravity theories. In
the parity-even gravity theory, we perform the $s$-mode stability
analysis by using the BTZ black string perturbations, which gives
two Schr\"odinger equations with frequency-dependent potentials. The
$s$-mode stability is consistent with the generalized
Breitenlohner-Freedman bound for spin-2 field.  It seems that  for
the parity-odd massive gravity theory, the BTZ black hole is stable
when the imaginary part of quasinormal frequencies of massive
graviton is negative. However, this condition is not consistent with
the $s$-mode stability based on the second-order equation obtained
after squaring the first-order equation. Finally, we explore the
black hole stability connection between the parity-odd and
parity-even massive gravity theories.

\end{abstract}
\vspace{5mm}

{\footnotesize ~~~~PACS numbers: 04.70.Bw }

\vspace{1.5cm}

\hspace{11.5cm}{Typeset Using \LaTeX}
\newpage
\renewcommand{\thefootnote}{\arabic{footnote}}
\setcounter{footnote}{0}

%%%% Introduction %%%%
\section{Introduction}
If  a black hole solution is known, it is very important to carry
out the stability analysis of the black hole. At the early stage of
studying the Schwarzschild black hole, a conventional method to
determine the stability is to solve the linearized Einstein equation
by choosing even-and odd-parity perturbations under the
Regge-Wheeler gauge for graviton, which leads to two Schr\"odinger
equations: Regge-Wheeler equation~\cite{Regge:1957td} and Zerilli
equation~\cite{Zerilli:1970se}. One may conclude that the
Schwarzschild black hole is stable because their potentials are
positive definite for the whole region outside the black hole,
implying that there is no exponentially growing
modes~\cite{Vishveshwara:1970cc,chan}. Equivalently, the stability
of a black hole depends on the sign of the imaginary part
$\omega_{\rm I}$  of their quasinormal frequencies
$\omega=\omega_{\rm R}+i \omega_{\rm I}$ when considering the time
dependence of $e^{-i\omega t}$~\cite{Berti:2009kk}. If $\omega_{\rm
I}$ is negative, the black hole is stable.  The real part
$\omega_{\rm R}$ has no bearing on stability properties.
Furthermore, the unstable condition of a black hole was suggested
to be $\omega_{\rm R}=0$ and $\omega_{\rm
I}\ge0$~\cite{Konoplya:2011qq}.

On the other hand, the stability analysis of the Schwarzschild
black hole obtained from  higher derivative gravity  is not an
easy task because it contains the second-order equation for a
massive graviton. A conventional stability method designed for a
graviton with two degrees of freedom (2 DOF) is not suitable for
studying the massive graviton (5 DOF)~\cite{Fierz:1939ix} which is
propagating on the black hole and de Sitter
spacetimes~\cite{Higuchi:1986py}. However, if one considers a
lower dimensional massive gravity, the situation is not so
complicated. Reminding that the three dimensional Einstein gravity
is a gauge theory, any propagating spin-2 mode belongs to massive
graviton which can be obtained from three-dimensional massive
gravity theories. Further, these theories are classified into
parity-even and parity-odd theories.

Recently, it was shown that the BTZ black
hole~\cite{Banados:1992wn,Banados:1992gq} is stable for all $\mu$
(Chern-Simons coupling constant) against the massive spin-2
perturbations in the topologically massive gravity
(TMG~\cite{Deser:1981wh}, parity-odd theory) by demanding
boundedness of the perturbation at the
horizon~\cite{Birmingham:2010gj}. On the other hand, it was
suggested that the BTZ black hole is stable for $m^2>1/2\ell^2$ in
new massive gravity (NMG, parity-even
theory)~\cite{Bergshoeff:2009hq} by computing quasinormal
frequencies and performing the $s$-mode
analysis~\cite{Myung:2011bn}.

Because of different parity, one uses  different stability analysis
for the BTZ black hole. Solving the first-order differential tensor
equation algebraically together with the boundary conditions, we
obtain all quasinormal frequencies of massive graviton for
parity-odd theories~\cite{Sachs:2008gt}: TMG  and generalized
massive gravity (GMG)~\cite{Bergshoeff:2009aq}.  If $\omega_{\rm I}
< 0$, the black hole seems to be  stable against the massive
graviton perturbation.

Given the parity-odd first order linearized equation, one obtains
its second-order linearized equation after squaring it, which
belongs to the parity-even theory, giving the ambiguity on sign of
the mass. Because of this ambiguity, someone prefers solving the
first-order equation directly~\cite{Sachs:2008gt}, instead of  the
second-order equation. Off-critical point, the parity-even gravity
theory usually provides the second-order linearized equation after
choosing the transverse-traceless gauge. It is known that ``solving
directly the second-order massive  equation" is a formidable task
for the BTZ black hole spacetimes. Fortunately, after choosing the
BTZ black string  perturbation for massive
graviton~\cite{Liu:2008ds}, the $s$-mode analysis may be performed
using two Schr\"odinger equations with frequency-dependent
potentials~\cite{Myung:2011bn}.

In this work, we study the massive graviton stability of the BTZ
black hole obtained from three-dimensional\footnote{It is well-known
that the  Einstein gravity in three dimensions has no propagating
degrees of freedom (DOF). This is  clearly shown  by counting a
massless graviton $h_{\mu\nu}$: $D(D-3)/2$. One has zero DOF for $D
=3$.  For a massive graviton,  it is changed into $ (D-2)(D + 1)/2$
which gives  2 DOF for $D = 3$. Thus,  massive generalizations of
the Einstein gravity~\cite{Bergshoeff:2009hq,Deser:1982vy}, allow
propagating degrees of freedom.  The three dimensional massive
gravity is regarded as a toy model of perturbative quantum gravity,
since we expect to have  less severe short-distance behavior than
four dimensional gravity with non-renormalizability.  We note that
if a black hole solution in three dimensional massive gravity is
found, a first issue is to examine its classical stability
properties as will be performed in the present paper.} massive
gravities which are classified into the parity-odd theories (TMG,
GMG) and parity-even gravity theories (NMG, six-derivative gravity
(SDG)~\cite{Bergshoeff:2012ev}). For the second-order
Schr\"{o}dinger  equation with an effective potential, we analyze
the massive graviton stability by checking if the potential is
positive for the whole range outside an event horizon.  However, we
should extend the stability condition  (\ref{flat-st}) for
asymptotically flat spacetimes to (\ref{xi-c}) for asymptotically
AdS spacetimes. In this case, the stability condition of $s$-mode is
given by  the generalized Breitenlohner-Freedman bound for spin-2
field. On the other hand, for a first-order massive graviton
equation, we perform the stability analysis by using the quasinormal
frequencies $(\omega=\omega_{\rm R}+i \omega_{\rm I})$. If
$\omega_{\rm I}$ is negative when considering the time dependence of
$e^{-i\omega t}$~\cite{Berti:2009kk}, the black hole is stable.
 Finally, we explore the black hole
stability connection between the parity-odd and parity-even
($s$-mode) massive gravity theories.

%%%%%%%%%%%%%%%%%%%%%%%%%%%%%%%%%%%%%%%%%%%%%%%%%%%%%%%%%%%%%%%%%%%%
%%%%%%%%%%%%%%%%%%%%%%%%%%%%%%%%%%%%%%%%%%%%%%%%%%%%%%%%%%%%%%%%%%%%
\section{Parity-even massive gravities}
%%%%%%%%%%%%%%%%%%%%%%%%%%%%%%%%%%%%%%%%%%%%%%%%%%%%%%%%%%%%%%%%%%%%
%%%%%%%%%%%%%%%%%%%%%%%%%%%%%%%%%%%%%%%%%%%%%%%%%%%%%%%%%%%%%%%%%%%%
In this work, we consider  the (non-rotating)
  BTZ black hole solution \cite{Banados:1992wn} which is a solution to all massive gravity theories. Its line element is
 given by
\begin{eqnarray}
ds_{{\rm BTZ}}^2&=&\bar{g}_{\mu\nu}dx^{\mu}dx^{\nu}\nn\\
 &=&-\left(-{\cal M}+\frac{r^2}{\ell^2}\right)dt^2+
 \left(-{\cal
 M}+\frac{r^2}{\ell^2}\right)^{-1}dr^2+r^2d\phi^2,\label{btz}
\end{eqnarray}
where ${\cal M}$ is the ADM mass given to be ${\cal M}=r_+^2/\ell^2$
with the horizon radius $r_+$ and AdS$_3$ curvature radius $\ell$.
Throughout the paper, the overbar denotes the background metric
\eqref{btz} for the BTZ black hole. The Ricci scalar, Ricci tensor,
Riemann tensor can be written in terms of the background metric
\eqref{btz} as follows:
\begin{eqnarray}
\bar{R}=6\Lambda,~~~~~\bar{R}_{\mu\nu}=2\Lambda
\bar{g}_{\mu\nu},~~~~~
\bar{R}_{\mu\nu\rho\sigma}=\Lambda(\bar{g}_{\mu\rho}\bar{g}_{\nu\sigma}
-\bar{g}_{\mu\sigma}\bar{g}_{\nu\rho}),
\end{eqnarray}
where $\Lambda=-1/\ell^2$. Also we adopt a notation of $(-,+,+)$ and
unit of $2\kappa^2=1$. For the perturbation around the BTZ black
hole
\begin{eqnarray}\label{pert}
g_{\mu\nu}=\bar{g}_{\mu\nu}+h_{\mu\nu},
\end{eqnarray} the
linearized Ricci tensor and Ricci scalar are given by
\begin{eqnarray}
\delta
R_{\mu\nu}(h)&=&\frac{1}{2}\Big(\bar{\nabla}_{\mu}\bar{\nabla}^{\rho}h_{\rho\nu}
+\bar{\nabla}_{\nu}\bar{\nabla}^{\rho}h_{\rho\mu}-\bar{\nabla}^2h_{\mu\nu}
-\bar{\nabla}_{\mu}\bar{\nabla}_{\nu}h\Big)+3\Lambda
h_{\mu\nu}-\Lambda \bar{g}_{\mu\nu}h,\label{lrmunu}\\
\delta
R(h)&=&\bar{\nabla}_{\alpha}\bar{\nabla}_{\beta}h^{\alpha\beta}-\bar{\nabla}^2h-2\Lambda
h\label{lr}.
\end{eqnarray}

Let us first introduce a three-dimensional massive gravity proposed
by Fierz and Pauli (FP) \cite{Fierz:1939ix}  whose action is given
by
\begin{eqnarray}\label{sfp}
S_{\rm FP}=S_{\rm bl}(h)-\frac{M_{\rm FP}^2}{4}\int
d^3x\sqrt{-g}\left(h_{\mu\nu}h^{\mu\nu}-h^2\right),
\end{eqnarray}
where $M_{\rm FP}^2$ is a mass parameter and $S_{\rm bl}(h)$ is the
bilinear form of the Einstein-Hilbert action with a cosmological
constant $\Lambda$. It is well known that the linearized Einstein
equation can be written as \cite{Higuchi:1986py}
\begin{eqnarray}\label{fpmeq}
\Big(\bar{\nabla}^{2}-2\Lambda-M_{\rm FP}^2\Big)h_{\mu\nu}^{\rm
FP}=0,
\end{eqnarray}
which is considered as a simplest equation for a massive graviton
with 2 DOF on the BTZ black hole spacetimes. In deriving this
equation, we have used the consistency condition of the linearized
Bianchi identity
\begin{equation} \label{tt-g} \bar{\nabla}_\mu
h^{\mu\nu}=0,~~h^{\mu}_{~\mu}=0,
\end{equation}
which is considered  as the transverse-traceless (TT)
gauge\footnote{We note that the action $S_{\rm FP}$ (\ref{sfp})
has no diffeomorphism invariance, while the TT gauge condition is
imposed only when considering diffeomorphism invariant actions of
$S_{\rm SDG},S_{\rm TMG}$, and $S_{\rm GMG}. $}.

 Now we  consider the parity-even
six-derivative gravity (SDG) \cite{Bergshoeff:2012ev}, whose action
is given by
\begin{eqnarray}\label{spet}
\hspace*{-1em}S_{{\rm SDG}}=\int d^3x\sqrt{-g} \Big[\sigma
R-2\lambda_S+\alpha R^2+\beta R_{\mu\nu}R^{\mu\nu}
+a_1\nabla_{\mu}R\nabla^{\mu}R+
a_2\nabla_{\rho}R_{\mu\nu}\nabla^{\rho}R^{\mu\nu}\Big],\label{pet}
\end{eqnarray}
where $\sigma=0,\pm1$ is a dimensionless parameter, $\lambda_S$ is a
cosmological parameter with mass dimension 2. Here parameters
$\alpha(\beta)$ have mass dimension $-2$ and $a_1(a_2)$ have $-4$.
We remark that when choosing $a_1=a_2=0$, $\sigma=1,$ and
$8\alpha+3\beta=0$, the action (\ref{spet}) reduces to the NMG
action \cite{Bergshoeff:2009hq} as
\begin{eqnarray}\label{snmg}
S_{\rm NMG}=\int d^3x\sqrt{-g}\Bigg[R-2\lambda_S-\frac{1}{m^2}\left(
R_{\mu\nu}R^{\mu\nu}-\frac{3}{8}R^2\right)\Bigg],
\end{eqnarray}
where $m^2$ is a mass parameter with dimension $2$.

Varying the action  \eqref{pet} with respect to  $g^{\mu\nu}$ leads
to the  equation
\begin{eqnarray}\label{peteq}
\sigma\Big(R_{\mu\nu}-\frac{1}{2}g_{\mu\nu}R\Big)+\lambda_S
g_{\mu\nu}+E_{\mu\nu}+H_{\mu\nu}=0
\end{eqnarray}
with
\begin{eqnarray}
E_{\mu\nu}&=&\beta\Big[-\frac{1}{2}g_{\mu\nu}R_{\rho\sigma}R^{\rho\sigma}
+2R_{\mu\rho\nu\sigma}R^{\rho\sigma}+\nabla_{\gamma}\nabla^{\gamma}
R_{\mu\nu}+\frac{1}{2}g_{\mu\nu}\nabla_{\gamma}\nabla^{\gamma}R
-\nabla_{\mu}\nabla_{\nu}R\Big]\nn\\
&&+\alpha\Big[2RG_{\mu\nu}+2g_{\mu\nu}\nabla_{\gamma}\nabla^{\gamma}R
-2\nabla_{\mu}\nabla_{\nu}R\Big],\\
&&\nn\\
H_{\mu\nu}&=&a_1\Big[\nabla_{\mu}R\nabla_{\nu}R-2R_{\mu\nu}\nabla^2R
-\frac{1}{2}g_{\mu\nu}\nabla_{\rho}R\nabla^{\rho}R-2(g_{\mu\nu}\nabla^2
-\nabla_{\mu}\nabla_{\nu}\nabla^2)R\Big]\nn\\
&&a_2\Big[\nabla_{\mu}R_{\rho\sigma}\nabla_{\nu}R^{\rho\sigma}
-\frac{1}{2}g_{\mu\nu}\nabla_{\gamma}R_{\rho\sigma}\nabla^{\gamma}R^{\rho\sigma}
-\nabla^2R_{\mu\nu}-g_{\mu\nu}\nabla^{\rho}\nabla^{\sigma}\nabla^2R_{\rho\sigma}
\nn\\
&&+2\nabla^{\rho}\nabla_{(\mu}\nabla^2R_{\nu)\rho}
+2\nabla^{\rho}R_{\rho\sigma}\nabla_{(\mu}R_{\nu)}^{\sigma}
+2R_{\rho\sigma}\nabla^{\rho}\nabla_{(\mu}R_{\nu)}^{\sigma}
-2R_{\sigma(\mu}\nabla^2R_{\nu)}^{\sigma}\nn\\
&&-2\nabla_{\rho}R_{\sigma(\mu}\nabla_{\nu)}R^{\rho\sigma}
-2R_{\sigma(\mu}\nabla^{\rho}\nabla_{\nu)}R_{\rho}^{\sigma}\Big].
\end{eqnarray}
We note that the BTZ black hole solution (\ref{btz}) to Eq.
(\ref{peteq}) is allowed only when choosing
$\lambda_{S}=\sigma\Lambda-2(3\alpha+\beta)\Lambda^2$. Taking into
account the perturbation (\ref{pert}) and plugging  the TT gauge
(\ref{tt-g}) into  Eq.(\ref{peteq}), we obtain  the sixth-order
differential perturbation equation, which can be factored into three
pieces\footnote{In order to eliminate  scalar graviton, we require
three conditions  as \cite{Bergshoeff:2012ev}
\begin{eqnarray} a_1=-3a_2/8,~~\alpha=\Lambda
a_2/8-3\beta/8,~~-\sigma/2+3\Lambda^2
a_2/4-\Lambda\beta/4\neq0.\nn
\end{eqnarray}
}:
\begin{eqnarray}\label{petpeq}
\Big[\bar{\nabla}^2-2\Lambda\Big]
\Big[\bar{\nabla}^2-2\Lambda-M_+^2\Big]
\Big[\bar{\nabla}^2-2\Lambda-M_-^2\Big]h_{\mu\nu}=0.
\end{eqnarray}
Here the mass parameters $M^2_{\pm}$ denote
\begin{eqnarray}
M_{\pm}^2=\frac{\beta}{2a_2}-\Lambda\pm\frac{1}{2a_2}
\sqrt{10a_2^2\Lambda^2-6a_2\beta\Lambda+4a_2\sigma+\beta^2}.
\end{eqnarray}
In  Eq. (\ref{petpeq}), we read off two massive equations
\begin{eqnarray}\label{petmeq}
\Big[\bar{\nabla}^2-2\Lambda-M_+^2\Big]h_{\mu\nu}^{M_+}=0,~~~
\Big[\bar{\nabla}^2-2\Lambda-M_-^2\Big]h_{\mu\nu}^{M_-}=0
\end{eqnarray}
off-critical points ($M^2_+\not=M^2_-$). They describe 4 DOF for two
massive gravitons.

 Also, we note that for the NMG (\ref{snmg}), the
linearized equation is given by
\begin{eqnarray}\label{nmgmeq}
\Big[\bar{\nabla}^2-2\Lambda -M^2_{\rm NMG}\Big]h_{\mu\nu}^{\rm
NMG}=0, ~~M_{\rm NMG}^2~=~m^2-\frac{1}{2\ell^2}
\end{eqnarray}
off critical point ($m^2\neq1/2\ell^2$) and off-decoupling
limit\footnote{In the decoupling limit of $m^2\to0$, however, NMG
action (\ref{snmg}) reduces to massless NMG \cite{Deser:2009hb}
where  the fourth order equation appears, instead of the  second
order equation.} ($m^2\neq0$), which describes 2 DOF for a massive
graviton in three dimensional spacetimes.
%%%%%%%%%%%%%%%%%%%%%%%%%%%%%%%%%%%%%%%%%%%%%%%%%%%%%%%%%%%%%%%%%%%%
%%%%%%%%%%%%%%%%%%%%%%%%%%%%%%%%%%%%%%%%%%%%%%%%%%%%%%%%%%%%%%%%%%%%
\subsection{$s$-mode stability analysis}
%%%%%%%%%%%%%%%%%%%%%%%%%%%%%%%%%%%%%%%%%%%%%%%%%%%%%%%%%%%%%%%%%%%%
%%%%%%%%%%%%%%%%%%%%%%%%%%%%%%%%%%%%%%%%%%%%%%%%%%%%%%%%%%%%%%%%%%%%

We are now in a position to perform the stability analysis of
massive gravitons satisfying Eqs. (\ref{fpmeq}), (\ref{petmeq}), and
(\ref{nmgmeq}).  We propose that they  are propagating  on the BTZ
black hole background (\ref{btz}).   For this purpose, inspired by
the BTZ black string perturbations~\cite{Liu:2008ds},  we consider
the following two distinct (orthogonal) perturbations ansatz
\cite{Myung:2011bn}:\\
 the
 type I has two off-diagonal components $h_0$ and $h_1$
\begin{eqnarray}
h^I_{\mu\nu}=\left(
\begin{array}{ccc}
0 & 0 & h_0(r) \cr 0 & 0 & h_1(r) \cr h_0(r) & h_1(r) & 0
\end{array}
\right) e^{-i\omega t}e^{ik\phi} \,, \label{type1}
\end{eqnarray}
while for the type II, the metric tensor takes the form with four
components $H_0,~H_1,~H_2,$ and $H_3$ as
\begin{eqnarray}
h^{II}_{\mu\nu}=\left(
\begin{array}{ccc}
H_0(r)  & H_1(r) & 0  \cr H_1(r) & H_2(r) & 0 \cr 0 & 0 & H_3(r)
\end{array}
\right) e^{-i\omega t}e^{ik\phi} \,. \label{type2}
\end{eqnarray}
In this work, we focus on $s$-mode ($k=0$) case for simplicity.
Importantly,   Eqs. (\ref{fpmeq}), (\ref{nmgmeq}), and
(\ref{petmeq}) can be combined into a single massive equation
\begin{eqnarray}\label{meq}
\left(\bar{\nabla}^2-2\Lambda-M_{i}^2\right)h_{\mu\nu}^{M_i}=0,
\end{eqnarray}
where  $M_i^2(i=1,2,3,4)$ is  given by
\begin{eqnarray}
M_1^2=M_{{\rm FP}}^2,~ M_2^2=M_{\rm
NMG}^2,~M_3^2=M_{+}^2,~M_4^2=M_{-}^2.
\end{eqnarray}
For type I, plugging (\ref{type1}) into (\ref{meq}) and eliminating
$h_1(r)$ from $(t,\phi)$ and $(r,\phi)$ components of (\ref{meq})
lead to the Schr\"{o}dinger equation as
\begin{eqnarray}\label{schPhi}
\frac{d^2\Phi_{i}}{dr^{*2}}+\Big[\omega^2-V^{\rm
I}_{\Phi_i}\Big]\Phi_{i}=0,
\end{eqnarray}
where $r^*$ is the tortoise coordinate defined by the relation of
$dr^*=\ell^2dr/(r^2-r_+^2)$. Here, a new field $\Phi_i$ is defined
by $\Phi_i=h_0/\sqrt{r\{M_i^2(r^2-r_+^2)/\ell^2-\omega^2\}}$, and
$V^{\rm I}_{\Phi_i}$ is the $\omega$-dependent potential given by
\begin{eqnarray}\label{v1}
&&V^{\rm I}_{\Phi_i}(\omega,r)=\frac{r^2-r_+^2}{\ell^2}
\Bigg[M_{i}^2+\frac{15}{4\ell^2}-\frac{3r_+^2}{4\ell^2r^2}+
\frac{3M_i^4 r^2(r^2-r_+^2)} {\ell^6\left\{M_i^2(r^2-r_+^2)/\ell^2-\omega^2\right\}^2}\nonumber\\
&&\hspace{11em}
+\frac{2M_i^2(2r_+^2-3r^2)}{\ell^4\left\{M_i^2(r^2-r_+^2)/\ell^2-\omega^2\right\}}\Bigg].
\end{eqnarray}
We show that for $M_i^2\ge0$, all potentials $V^{\rm I}_{\Phi_i}$
are always positive definite for the whole range of $r_+\le
r\le\infty$. This may imply that for  $M_i^2\ge0$,  the BTZ black
hole  is stable against type I perturbation.

On the other hand, in type II case, substituting (\ref{type2}) into
(\ref{meq}) and after some manipulations, $(t,\phi)$ component of
(\ref{meq}) can be written as the other  Schr\"{o}dinger equation:
\begin{eqnarray} \label{schpsi}
\frac{d^2\Psi_i}{dr^{*2}}+[\omega^2-V^{\rm
II}_{\Psi_i}(\omega,r)]\Psi_i=0,
\end{eqnarray}
where $\Psi_i=H_1\sqrt{r(r^2-r_+^2)^2}/
\sqrt{M_i^2(r_+^2-r^2)\ell^2+(2r_+^2-r^2)+\omega^2\ell^4}$ and the
$\omega$-dependent potential $V^{\rm II}_{\Psi_i}$ is given by
\begin{eqnarray}\label{v2}
&&\hspace{-2em}V^{\rm
II}_{\Psi_i}(\omega,r)=\frac{r^2-r_+^2}{\ell^2}
\Bigg[M_i^2+\frac{7}{4\ell^2}-\frac{3r_+^2}{4\ell^2r^2}
 +\frac{3r^2(M_i^2+1/\ell^2)^2(r^2-r_+^2)} {\ell^6
 \left\{M_i^2(r_+^2-r^2)/\ell^2+(2r_+^2-r^2)/\ell^4+\omega^2\right\}^2}
\nonumber\\&&\hspace{8em}+\frac{4(M_i^2+1/\ell^2)(r^2-r_+^2)}
{\ell^4\left
\{M_i^2(r_+^2-r^2)/\ell^2+(2r_+^2-r^2)/\ell^4+\omega^2\right\}}\Bigg].
\end{eqnarray}
We note that for $M_i^2\ge0$, all  potential $V^{\rm II}_{\Psi_i}$
is always positive definite for the whole range of $r_+\le
r\le\infty$, which states  that the BTZ black hole is stable against
type-II perturbation.

Hence, if one applies  type I and II perturbations to the
parity-even massive gravities, the stability conditions of the BTZ
black hole seem to be
\begin{eqnarray} \label{flat-st}
M_{\rm FP}^2 \ge 0,~~~~ m^2\ge \frac{1}{2\ell^2}, ~~~~~
M_{\pm}^2\ge0
\end{eqnarray}
in FP, NMG, and SDG, respectively. However, these conditions are
suitable for asymptotically flat spacetimes. We remind the reader
that our spacetime is asymptotically anti de Sitter spacetimes.
Therefore, we have to point out what is  the stability condition of
a massive graviton propagating on the AdS$_3$ spacetimes.   To see
this explicitly, let us consider asymptotically AdS$_3$ spacetimes,
which corresponds to a large $r$ limit ($r^{*}\to0$) in
Eq.(\ref{btz}). In this limit, the potentials (\ref{v1}) and
(\ref{v2})  take the same form  when expressing them in terms of a
tortoise coordinate $r^*$
\begin{eqnarray}
V^{\rm I}_{\Phi_i},~V^{\rm
II}_{\Psi_i}\sim\frac{\xi}{r^{*2}},\end{eqnarray} where
\begin{eqnarray} \xi=\ell^2\left(M_i^2+\frac{3}{4\ell^2}\right).
\end{eqnarray}
 As $r^{*}$ approaches $0$, Eqs. (\ref{schPhi}) and (\ref{schpsi})
become one-dimensional Schr\"{o}dinger equation with an inverse
square potential of the strength $\xi$ and the energy $E=\omega^2$.
It is known~\cite{Case:1950an,Moon:2012dy} that in this case, if
$\xi$ satisfies the condition,
\begin{eqnarray} \label{xi-c}
\xi\ge-\frac{1}{4}~~\Rightarrow~~M_i^2\ge-\frac{1}{\ell^2},
\end{eqnarray}
the energy spectrum is always continuous and positive. It is worth
noting that the stability condition (\ref{xi-c}) is consistent with
the regularized condition at $r^{*}=0$. Importantly, the stability
condition (\ref{xi-c}) is exactly the same with the
Breitenlohner-Freedman (BF) bound \cite{Breitenlohner:1982bm} for a
massive spin-2 field in AdS$_3$
spacetimes~\cite{Gover:2008sw,Lu:2011qx}
\begin{eqnarray}
\hspace*{-2em}\Big[\nabla_{(\rm AdS)}^2-2\Lambda-M_{(\rm
AdS)}^2\Big]h_{\mu\nu}=0~~~\Rightarrow~~~M_{(\rm AdS)}^2\ge M_{\rm
BF}^2=-\frac{1}{\ell^2}.
\end{eqnarray}
Hence, we should extend the stability condition (\ref{flat-st}) for
asymptotically flat spacetimes to the stability condition
(\ref{xi-c}) for asymptotically AdS spacetimes.

Finally, we dictates the stability condition of the BTZ black hole
\begin{equation} \label{e-sc}
M_{\rm FP}^2\ge-\frac{1}{\ell^2},~~~~ m^2\ge -\frac{1}{2\ell^2},
~~~~{\rm and}~~ M_{\pm}^2\ge -\frac{1}{\ell^2}
\end{equation}
off-critical points ($m^2\neq1/2\ell^2,~M_{\pm}^2\neq0$) and
off-decoupling limit ($m^2\neq0$), when using the $s$-mode analysis
for the parity-even massive gravity theories.

%%%%%%%%%%%%%%%%%%%%%%%%%%%%%%%%%%%%%%%%%%%%%%%%%%%%%%%%%%%%%%%%%%%%
%%%%%%%%%%%%%%%%%%%%%%%%%%%%%%%%%%%%%%%%%%%%%%%%%%%%%%%%%%%%%%%%%%%%
\section{Parity-odd massive gravities}
%%%%%%%%%%%%%%%%%%%%%%%%%%%%%%%%%%%%%%%%%%%%%%%%%%%%%%%%%%%%%%%%%%%%
%%%%%%%%%%%%%%%%%%%%%%%%%%%%%%%%%%%%%%%%%%%%%%%%%%%%%%%%%%%%%%%%%%%%
A parity-odd massive gravity in three dimensions was first
introduced  by Deser, Jackiw, and Templeton \cite{Deser:1982vy}. The
TMG  action includes a gravitational Chern-Simons term, which
reveals parity-violation or `odd' parity. In this section, we
introduce two parity-odd massive gravities of TMG and GMG, and
investigate the stability analysis of the BTZ black hole in those
gravities.
%%%%%%%%%%%%%%%%%%%%%%%%%%%%%%%%%%%%%%%%%%%%%%%%%%%%%%%%%%%%%%%%%%%%
%%%%%%%%%%%%%%%%%%%%%%%%%%%%%%%%%%%%%%%%%%%%%%%%%%%%%%%%%%%%%%%%%%%%
\subsection{TMG}
%%%%%%%%%%%%%%%%%%%%%%%%%%%%%%%%%%%%%%%%%%%%%%%%%%%%%%%%%%%%%%%%%%%%
%%%%%%%%%%%%%%%%%%%%%%%%%%%%%%%%%%%%%%%%%%%%%%%%%%%%%%%%%%%%%%%%%%%%
The action of TMG  with a negative
cosmological constant is given by \cite{Deser:1982vy}
\begin{eqnarray}
S_{{\rm TMG}}=\int d^3x\sqrt{-g}\Big[R-2\Lambda +\frac{1}{2\mu}
\epsilon^{\lambda\mu\nu}\Gamma_{\lambda\sigma}^{\rho}
\Big(\partial_{\mu}\Gamma_{\rho\nu}^{\sigma}
+\frac{2}{3}\Gamma_{\mu\tau}^{\sigma}\Gamma_{\nu\rho}^{\tau}\Big)\Big],
\end{eqnarray}
where $\mu$ is a parameter with mass dimension 1. The Einstein
equation takes the form
\begin{eqnarray}\label{tmg}
R_{\mu\nu}-\frac{1}{2}g_{\mu\nu}R+\Lambda g_{\mu\nu}
+\frac{1}{\mu}C_{\mu\nu}=0,
\end{eqnarray}
where the Cotton tensor $C_{\mu\nu}$ is defined by
\begin{eqnarray}\label{cmunu}
C_{\mu\nu}\equiv\epsilon_{\mu}^{~\alpha\beta}\nabla_{\alpha}
(R_{\beta\nu}-\frac{1}{4}g_{\beta\nu}R).
\end{eqnarray}
Introducing the perturbation (\ref{btz})
and applying the TT gauge condition (\ref{tt-g}) to the linearized
equation of $(\ref{tmg})$, we arrive at
\begin{eqnarray}\label{tmgpeq}
\Big[\bar{\nabla}^2-2\Lambda\Big]\Big[h_{\mu\nu} +\frac{1}{\mu}
\epsilon_{\mu}^{~\alpha\beta}\bar{\nabla}_{\alpha}h_{\beta\nu}\Big]=0.
\end{eqnarray}
From (\ref{tmgpeq}), we read off the first-order differential equation for a
massive graviton
\begin{eqnarray}\label{tmgeq}
\epsilon_{\mu}^{~\alpha\beta}\bar{\nabla}_{\alpha}h_{\beta\nu} +\mu
h_{\mu\nu} =0.
\end{eqnarray}
One can easily check that squaring it [equivalently, by applying
the first-order operator
$\epsilon_{\sigma}^{~\rho\mu}\bar{\nabla}_\rho-\mu
\delta_{\sigma}^{\mu}$ to (\ref{tmgeq})] leads to the second-order
equation
\begin{eqnarray}\label{tmgeq2}
\Big[\bar{\nabla}^2-2\Lambda-M_{\rm TMG}^2\Big]h_{\sigma\nu}=0
\end{eqnarray}
with $M_{\rm TMG}^2=\mu^2-1/\ell^2$. Using the bound given by the
stable condition (\ref{e-sc}), we have
\begin{equation}
M_{\rm TMG}^2 \ge -\frac{1}{\ell^2} \to \mu^2 \ge 0 \to |\mu| \ge 0
\end{equation}
which is consistent with that obtained in~\cite{Birmingham:2010gj},
indicating that the BTZ black hole is stable for all $\mu$ against
the massive spin-2 perturbation in TMG by demanding boundedness of
the perturbation at the horizon. The latter condition eliminates
 modes  which are growing in time and obeying the generalized boundary
conditions at asymptotic infinity. At this stage, we emphasize that
the authors in~\cite{Birmingham:2010gj} have used not the
first-order equation (\ref{tmgeq}) itself but a second-order
hypergeometric equation obtained by transforming two first-order
equations when analyzing the stability of the BTZ black hole.

%%%%%%%%%%%%%%%%%%%%%%%%%%%%%%%%%%%%%%%%%%%%%%%%%%%%%%%%%%%%%%%%%%%%
%%%%%%%%%%%%%%%%%%%%%%%%%%%%%%%%%%%%%%%%%%%%%%%%%%%%%%%%%%%%%%%%%%%%
\subsection{GMG}
%%%%%%%%%%%%%%%%%%%%%%%%%%%%%%%%%%%%%%%%%%%%%%%%%%%%%%%%%%%%%%%%%%%%
%%%%%%%%%%%%%%%%%%%%%%%%%%%%%%%%%%%%%%%%%%%%%%%%%%%%%%%%%%%%%%%%%%%%
We consider the GMG action which
consists of NMG and gravitational Chern-Simons term as \cite{Bergshoeff:2009hq,Liu:2009pha}
\begin{eqnarray}
&&\hspace*{-2em}S_{{\rm GMG}}=\int d^3x\sqrt{-g}\Big[\sigma
R-2\lambda_G +\frac{1}{m^2}\left(
R_{\mu\nu}R^{\mu\nu}-\frac{3}{8}R^2\right)+\frac{1}{2\mu}
\epsilon^{\lambda\mu\nu}\Gamma_{\lambda\sigma}^{\rho}
\Big(\partial_{\mu}\Gamma_{\rho\nu}^{\sigma}
+\frac{2}{3}\Gamma_{\mu\tau}^{\sigma}\Gamma_{\nu\rho}^{\tau}\Big)\Big],
\nn\\
&&\label{sgmg}
\end{eqnarray}
where $\lambda_G$ is a cosmological parameter with mass dimension 2.
From the GMG action, one derives  the Einstein equation
\begin{eqnarray}\label{gmgeq}
\sigma G_{\mu\nu}+\lambda_G
g_{\mu\nu}+\frac{1}{2m^2}K_{\mu\nu}+\frac{1}{\mu}C_{\mu\nu}=0,
\end{eqnarray}
where $C_{\mu\nu}$ is given by Eq.(\ref{cmunu}) and $K_{\mu\nu}$
takes the form
\begin{eqnarray}\label{kmunu}
  K_{\mu\nu}&=&2\nabla^2R_{\mu\nu}-\frac{1}{2}\nabla_\mu \nabla_\nu R-\frac{1}{2}\nabla^2Rg_{\mu\nu}\nonumber\\
        &+&4R_{\mu\rho\nu\sigma}R^{\rho\sigma} -\frac{3}{2} R R_{\mu\nu}-R_{\rho\sigma}R^{\rho\sigma}g_{\mu\nu}
         +\frac{3}{8}{R}^2 g_{\mu\nu}.
\end{eqnarray} It is pointed out that the BTZ black
hole solution (\ref{btz}) is allowed only for
$\lambda_G=\Lambda^2/4m^2+\sigma\Lambda$. Using (\ref{pert}) and the
TT gauge condition (\ref{tt-g}), the linearized equation of
(\ref{gmgeq}) can be written as
\begin{eqnarray}
\Big[\bar{\nabla}^2-2\Lambda\Big] \Big[\bar{\nabla}^2h_{\mu\nu}
+\frac{m^2}{\mu}\epsilon_{\mu}^{~\alpha\beta}\bar{\nabla}_{\alpha}
h_{\beta\nu}+\left(\sigma
m^2-\frac{5}{2}\Lambda\right)h_{\mu\nu}\Big]=0.
\end{eqnarray}
Considering the above equation, we read off the second-order equation of the massive
graviton
\begin{eqnarray}\label{gmgpeq}
\bar{\nabla}^2h_{\mu\nu}
+\frac{m^2}{\mu}\epsilon_{\mu}^{~\alpha\beta}\bar{\nabla}_{\alpha}
h_{\beta\nu}+\left(\sigma m^2-\frac{5}{2}\Lambda\right)h_{\mu\nu}=0
\end{eqnarray}
which is further factorized into
\begin{eqnarray}
\Big[\delta_{\mu}^{\beta}+
\frac{1}{m_+}\epsilon_{\mu}^{~\rho\beta}\bar{\nabla}_{\rho}
\Big]\Big[\delta_{\beta}^{\gamma}
+\frac{1}{m_-}\epsilon_{\beta}^{~\sigma\gamma}\bar{\nabla}_{\sigma}
\Big]h_{\gamma\nu}=0.
\end{eqnarray}
Here $m_{\pm}$ take the forms
\begin{eqnarray}\label{mpm}
m_{\pm}=\frac{m^2}{2\mu}\pm\sqrt{\frac{m^4}{4\mu^2}-\sigma
m^2-\frac{\Lambda}{2}}.
\end{eqnarray}
This  implies that two massive gravitons with mass $m_\pm$  are
described by two first-order equations, respectively,
\begin{eqnarray}\label{gmgpm}
\epsilon_{\mu}^{~\alpha\beta}\bar{\nabla}_{\alpha}h_{\beta\nu} +m_+
h_{\mu\nu}
=0,~~~\epsilon_{\mu}^{~\alpha\beta}\bar{\nabla}_{\alpha}h_{\beta\nu}
+m_- h_{\mu\nu} =0.
\end{eqnarray}
As squaring their first-order equations, acting two operations
[$\epsilon_{\sigma}^{~\rho\mu}
\bar{\nabla}_{\rho}-m_+\delta_{\sigma}^{\mu}$] and
[$\epsilon_{\sigma}^{~\rho\mu}
\bar{\nabla}_{\rho}-m_-\delta_{\sigma}^{\mu}$] on (\ref{gmgpm})
leads to two second-order equations
\begin{eqnarray}\label{gmgpm2}
\Big[\bar{\nabla}^2-2\Lambda-M_{\rm
GMG+}^2\Big]h_{\sigma\nu}=0,~~~\Big[\bar{\nabla}^2-2\Lambda-M_{\rm
GMG-}^2\Big]h_{\sigma\nu}=0,
\end{eqnarray}
where
\begin{equation}
M_{{\rm GMG}\pm}^2=m_{\pm}^2-\frac{1}{\ell^2}.
\end{equation}
 Using the bound given by the stable condition (\ref{e-sc}),
we have the mass bound
\begin{equation} \label{even-st}
M_{\rm GMG \pm}^2 \ge -\frac{1}{\ell^2} \to m_{\pm}^2 \ge 0.
\end{equation}

Consequently,  the $s$-mode stability condition  after squaring
their first-order equations is given by
\begin{equation} \label{s-sodd}
m^2_i \ge 0,
\end{equation}
where
\begin{equation}
m^2_1=\mu^2({\rm TMG}),~~~~m^2_2=m^2_+({\rm
GMG}),~~~~m^2_3=m^2_-({\rm GMG}).
\end{equation}
%%%%%%%%%%%%%%%%%%%%%%%%%%%%%%%%%%%%%%%%%%%%%%%%%%%%%%%%%%%%%%%%%%%%
%%%%%%%%%%%%%%%%%%%%%%%%%%%%%%%%%%%%%%%%%%%%%%%%%%%%%%%%%%%%%%%%%%%%
\subsection{Quasinormal mode analysis}
%%%%%%%%%%%%%%%%%%%%%%%%%%%%%%%%%%%%%%%%%%%%%%%%%%%%%%%%%%%%%%%%%%%%
%%%%%%%%%%%%%%%%%%%%%%%%%%%%%%%%%%%%%%%%%%%%%%%%%%%%%%%%%%%%%%%%%%%%
We note that (\ref{tmgeq}) and (\ref{gmgpm}) belong to the
first-order equation and they are parity-odd, while (\ref{tmgeq2})
and (\ref{gmgpm2}) are the second-order equation and are
parity-even. Furthermore, (\ref{tmgeq2}) and (\ref{gmgpm2}) have
ambiguities on the sign of mass. Hence, it would be better to use
(\ref{tmgeq}) and (\ref{gmgpm}) than (\ref{tmgeq2}) and
(\ref{gmgpm2}) when considering an another stability analysis for
the parity-odd theories.

In this section, we redo  the stability analysis of the massive
graviton for TMG and GMG by computing quasinormal frequencies. For
this purpose, we note that  the type I and II perturbations are
suitable for the second-order differential equations (\ref{tmgeq2})
and (\ref{gmgpm2}), while these are inappropriate for applying to
the first-order equations (\ref{tmgeq}) and  (\ref{gmgpm}) directly.
As will be shown in Appendix,  applying type I and II to
(\ref{tmgeq}) and (\ref{gmgpm}) leads to all null perturbations due
to the parity-oddness of their equations.

Therefore, we perform  the stability analysis with solving
(\ref{tmgeq}) and (\ref{gmgpm})
 to find quasinormal frequencies by following the approach developed in \cite{Sachs:2008gt}.
We first note  that (\ref{tmgeq}) and  (\ref{gmgpm}) can be
written as a single first-order equation
\begin{eqnarray}\label{first}
\epsilon_{\mu}^{~\alpha\beta}\bar{\nabla}_{\alpha}h_{\beta\nu} +m_i
h_{\mu\nu} =0,
\end{eqnarray}
where $m_i$ denote $\mu$ and $m_{\pm}$. For our purpose, we consider
a BTZ black hole metric with ${\cal M}=1$ and $\ell=1$ $ (r_+=1)$
given in global coordinates
\begin{eqnarray}
ds^2_{\rm gc}=\bar{g}_{\mu\nu}dx^{\mu}dx^{\nu}=-\sinh^2\rho dt^2+\cosh^2\rho
d\phi^2+d\rho^2,\label{btz1}
\end{eqnarray}
which is obtained by  replacing $r$ by $r=\cosh\rho$ in (\ref{btz}).
In what follows we use light-cone coordinates with $u=t+\phi$ and
$v=t-\phi$ as
\begin{eqnarray}\label{lc}
ds^2_{\rm
lc}=\bar{g}_{\mu\nu}dx^{\mu}dx^{\nu}=\frac{1}{4}du^2+\frac{1}{4}dv^2
-\frac{1}{2}\cosh2\rho dudv+d\rho^2.
\end{eqnarray}
The  metric (\ref{lc}) admits isometry group of
SL(2,R)$\times$SL(2,R), which allows us two sets of the Killing
vector fields, $L_{0,\pm1}$ and $\bar{L}_{0,\pm1}$ given as
\begin{eqnarray}
L_0&=&-\partial_u,\nn\\
L_{-1}&=&e^{-u}\left(-\frac{\cosh2\rho}{\sinh2\rho}\partial_u
-\frac{1}{\sinh2\rho}\partial_v-\frac{1}{2}\partial_{\rho}\right),\nn\\
L_{1}&=&e^{u}\left(-\frac{\cosh2\rho}{\sinh2\rho}\partial_u
-\frac{1}{\sinh2\rho}\partial_v+\frac{1}{2}\partial_{\rho}\right)
\label{L}
\end{eqnarray}
and $\bar{L}_{0,\pm1}$ are defined by
operation of  $u\leftrightarrow v$ in (\ref{L}). Three vector fields
$L_{0,\pm1}$ satisfy the SL(2,R) algebra
\begin{eqnarray}
[L_0,L_{\pm1}]=\mp L_{\pm1},~~~[L_1,L_{-1}]=2L_0.
\end{eqnarray}
Taking  the perturbation ansatz in terms of  $(u,v,\rho)$
\begin{eqnarray}
h_{\mu\nu}=e^{-i\omega t-ik\phi}\psi_{\mu\nu}(\rho) =e^{-ip_+ u-ip_-
v}\psi_{\mu\nu}(\rho),~~~p_{\pm}=\frac{1}{2}(\omega\pm
k),
\end{eqnarray}
the $s$-mode ($k=0$) solutions to the first-order equation
(\ref{first}) are  given by  right($r$)/left($l$) moving modes:
\begin{eqnarray}
h^{r}_{\mu\nu}=e^{-2h_r t}\psi_{\mu\nu}= e^{-2h_r
t}(\sinh\rho)^{-2h_r}\left(
\begin{array}{ccc}
1 & 0 & \frac{2}{\sinh2\rho} \cr 0 & 0 & 0 \cr \frac{2}{\sinh2\rho}
& 0 & \frac{4}{\sinh^22\rho}
\end{array}
\right)\,,~~~h_r=-\frac{1}{2}(m_i+1) \label{sol1}
\end{eqnarray}
for $p_-=-ih_r$ and
\begin{eqnarray}
h^{l}_{\mu\nu}=e^{-2h_lt}\psi_{\mu\nu}=
e^{-2h_lt}(\sinh\rho)^{-2h_l}\left(
\begin{array}{ccc}
0 & 0 & 0 \cr 0 & 1 & \frac{2}{\sinh2\rho} \cr 0 &
\frac{2}{\sinh2\rho} & \frac{4}{\sinh^22\rho}
\end{array}
\right)\,,~~~h_l=\frac{1}{2}(m_i-1) \label{sol2}
\end{eqnarray}
for $p_+=-ih_l$. Note that the solution (\ref{sol1}) satisfies the
chiral highest weight condition of $\bar{L}_1h_{\mu\nu}=0$ and
(\ref{sol2}) satisfies  the anti-chiral highest weight condition
of $L_1h_{\mu\nu}=0$ which are equivalent to  the transversality
condition of $\bar{\nabla}_{\mu}h^{\mu}_{~\nu}=0$. However,
requiring both conditions leads to null modes.  Considering
relations $p_\pm=\omega^{r/l}/2$, the corresponding quasinormal
frequencies, whose quasinormal modes satisfy the boundary
conditions: ingoing modes at the horizon and Dirichlet boundary
condition at infinity, can be written as
\begin{eqnarray}
&&\omega^r=-2ih_r=2i \Big(\frac{m_i}{2}+\frac{1}{2}\Big),\\
&&\omega^l=-2ih_l=2i \Big(-\frac{m_i}{2}+\frac{1}{2}\Big).
\end{eqnarray}
The complete tower of right- and left-moving quasinormal modes is
generated by acting $L_{-1}\bar{L}_{-1}$ on $h^{r/l}_{\mu\nu}$ $n$
times as
\begin{equation}
h^{(n),~ r/l}_{\mu\nu}
=\Big(L_{-1}\bar{L}_{-1}\big)^nh^{r/l}_{\mu\nu},
\end{equation}
which leads to their quasinormal frequencies with overtone number
$n$
\begin{eqnarray}
&&\omega^r_{n}=-2i(h_r+n),\\
&&\omega^l_{n}=-2i(h_l+n).
\end{eqnarray}

 Since the stability condition is determined  by two basic
quasinormal frequencies with $\omega^{r/l}_{\rm I}<0$ for
$\omega^{r/l}=\omega_{{\rm R}}^{r/l}+i\omega_{\rm I}^{r/l}$, it is
given by
\begin{eqnarray} \label{odd-s}
|m_i|>1/\ell,
\end{eqnarray}
where the AdS$_3$ curvature radius $\ell$ is restored for
convenience. It seems, however, that there is some discrepancy
between  (\ref{odd-s}) and (\ref{s-sodd}).

\section{Discussions}
In this work, we have established the stability of the massive
graviton around BTZ black hole in massive gravity theories which are
classified into the parity-even gravity theories (NMG, SDG) and the
parity-odd theories (TMG, GMG). For the parity-even massive
gravities, the stability conditions employed by the $s$-mode
analysis are exactly the same with the BF-bound, which corresponds
to $M_i^2\ge-1/\ell^2$ (\ref{xi-c}). For the parity-even gravity
theories, the $s$-mode analysis and the BF-bound based on the
second-order massive equation are consistent with the
Birmingham-Mokhtari-Sachs result requiring the boundedness of
perturbation at the horizon where a second-order hypergeometric
equation was used. These are given by the condition of $m^2_i \ge 0
 ~(|m_i|\ge0)$.

We stress  that  the stability analysis performed by the quasinormal
frequencies gave a condition of $|m_i|>1/\ell$, being different from
$|m_i|\ge0$.
 We may  interpret it by mentioning that the connection between potential and quasinormal frequencies
 condition is guaranteed  if  the second-order equation  is used  as  Schr\"odinger equation~\cite{Moon:2011sz,Berti:2009kk,Konoplya:2011qq}.
 We here have obtained the quasinormal frequencies by using the first-order equation.
Hence, the condition of $|m_i|>1/\ell$ based on quasinormal modes
does not comprise the stability condition $|m_i| \ge 0$  obtained by
solving the second-order equation.  In this case,  the unstable
quasinormal modes exist for $0\le|m_i|\le1/\ell$. It is, however,
pointed out that these unstable modes may  be truncated by requiring
the boundedness at the horizon when considering the generalized
boundary conditions at asymptotic infinity \cite{Birmingham:2010gj}.
Hence, it suggests that  the stability condition might   be extended
to comprise
\begin{eqnarray}
|m_i|\ge0.
\end{eqnarray}

Finally, we would like to mention that the stability of a black hole
in four-dimensional massive gravity is determined by the
Gregory-Laflamme instability of a five-dimensional  black string. It
turned out that the small Schwarzschild black hole in the dRGT
massive gravity~\cite{Babichev:2013una} and fourth-order
gravity~\cite{Myung:2013doa} is unstable against the metric and
Ricci tensor perturbations.  In the present work, the stability was
mainly determined by the asymptotes of black hole spacetimes. Hence,
it suggests for a future direction that the stability of the massive
graviton around a BTZ black hole  would be revisited   by using the
(in)stability of four-dimensional black string.

\section*{\bf Acknowledgments}

 This work was supported  by the National
Research Foundation of Korea (NRF) grant funded by the Korea
government (MEST) (No.2012-R1A1A2A10040499).

\section*{Appendix: Inappropriateness of type I and II for parity-odd theory}

We first substitute type I perturbation (\ref{type1}) into the TMG
equation (\ref{tmgeq}). It turns out that  $(t,\phi)$ and
$(r,\phi)$ components of  (\ref{tmgeq}) yield just $\mu h_0(r)=0$
and $\mu h_1(r)=0$, which implies that type I perturbation becomes
null unless $\mu=0$.  Similarly, for type II perturbation
(\ref{type2}),  the components $(t,t), (r,r), (\phi,r),$ and
$(\phi,\phi)$ of
 (\ref{tmgeq}) are given by
\begin{eqnarray}
\mu H_0(r)~=~0, ~~~\mu H_2(r)~=~0,~~~\mu H_1(r)&=&0,~~~\mu
H_3(r)~=~0,\nn
\end{eqnarray}
which leads to all null components for the type II perturbation of
 $H_0=H_1=H_2=H_3=0$ unless $\mu=0$.

For GMG,  applying type I perturbation (\ref{type1}) to
(\ref{gmgpeq}), we find that the corresponding solution to $(t,r)$
and $(r,r)$ components of Eq.(\ref{gmgpeq}) is given by
$h_0(r)=h_1(r)=0$.   In type II case (\ref{type2}), we note  that
$H_0(r)$, $H_2(r)$, and $H_3(r)$ can be expressed in terms of
$H_1(r)$ by considering the traceless condition, $(\phi,t)$
component, and $(\phi,r)$ component in (\ref{gmgpeq}), respectively:
\begin{eqnarray} \label{adx}
&&H_0(r)=\frac{{\cal M}-r^2/\ell^2}{\omega r}\Big[({\cal
M}-3r^2/\ell^2)H_1(r)+r({\cal M}-r^2/\ell^2)H_1^{\prime}(r)\Big],\nn\\
&&H_2(r)=\frac{1}{\omega({\cal M}-r^2/\ell^2)}\Big[({\cal
M}-r^2/\ell^2)H_1^{\prime}(r)-2rH_1(r)/\ell^2\Big],\nn \\
&&H_3(r)=-\frac{r({\cal M}-r^2/\ell^2)}{\omega}H_1(r).\nn\\
\end{eqnarray}
Substituting (\ref{adx}) into $(t,r)$ and $ (t,\phi)$ components
of (\ref{gmgpeq}), we find  $H_1(r)=0$. In this case, it yields
$H_0(r)=H_2(r)=H_3(r)=0$ when using (\ref{adx}) again. As a result
type II perturbation becomes null for parity-odd gravity theories.

This proves the inappropriateness of type I and II for parity-odd
theory.

\end{document}